\pdfoutput=1

\documentclass[
    ,final            
  ]
  {aipproc}

\layoutstyle{6x9}

\usepackage{graphicx} 
\usepackage{epstopdf}

\newcommand{\Fig}[1]{Figure~\ref{#1}}

\newcommand{\mf}[1]{\ensuremath{\mathcal{#1}}}		
\renewcommand{\_}[1]{\ensuremath{_\mathrm{#1}}} 	

\begin{document}

\title{Two Color Entanglement}

\classification{03.65.Ud, 03.67.Bg, 42.50.Lc, 42.65.Yj}
\keywords      {quadrature entanglement, quantum communication, quantum optics}

\author{Aiko Samblowski}{
  address={Institut f\"ur Gravitationsphysik, Leibniz Universit\"at Hannover and Max-Planck-Institut f\"ur Gravitationsphysik (Albert-Einstein-Institut), Callinstr.~38, 30167~Hannover, Germany}
}

\author{Christina E.~Lauk\"otter}{
  address={Institut f\"ur Gravitationsphysik, Leibniz Universit\"at Hannover and Max-Planck-Institut f\"ur Gravitationsphysik (Albert-Einstein-Institut), Callinstr.~38, 30167~Hannover, Germany}
}

\author{Nicolai Grosse}{
  address={Department of Quantum Science, Research School of Physics \& Engineering,
The Australian National University, Australian Capital Territory 0200, Australia}
}

\author{Ping Koy Lam}{
  address={Department of Quantum Science, Research School of Physics \& Engineering,
The Australian National University, Australian Capital Territory 0200, Australia}
}

\author{Roman Schnabel}{
  address={Institut f\"ur Gravitationsphysik, Leibniz Universit\"at Hannover and Max-Planck-Institut f\"ur Gravitationsphysik (Albert-Einstein-Institut), Callinstr.~38, 30167~Hannover, Germany} 
}

\begin{abstract}
We report on the generation of entangled states of light between the wavelengths 810 and 1550\,nm in the continuous variable regime. The fields were produced by type I optical parametric oscillation in a standing-wave cavity build around a periodically poled potassium titanyl phosphate crystal, operated above threshold. Balanced homodyne detection was used to detect the non-classical noise properties, while filter cavities provided the local oscillators by separating carrier fields from the entangled sidebands. We were able to obtain an inseparability of $\mf{I}=0.82$, corresponding to about $-0.86$\,dB of non-classical quadrature correlation. 
\end{abstract}

\maketitle


\section{Introduction}
Entangled states of light are the fundamental resource of many quantum communication and information protocols \cite{Braunstein2002}. Especially desirable for future quantum networks are quantum connections between atomic transitions and light that can be transmitted via telecom fibers. 
Fibers with an ultralow attenuation of 0.17\,dB/km at 1550\,nm are commercially available today \cite{Li2008}, while quantum states can be stored in alkaline atoms \cite{Honda2008,Appel2008} that absorb and emit light at wavelengths around 810\,nm \cite{Sansonetti2009}.
%
%
Recently, Li \emph{et~al.} \cite{Li2010} demonstrated the generation of bright entangled twin beams at these wavelength, using single ended filter cavities to detect the entanglement \cite{Coelho2009}. 

In this paper we present the experimental realization of a continuous wave (CW) source of entanglement between 810 and 1550\,nm in the continuous variable (CV) regime. A non-degenerate optical parametric oscillator (NOPO) was used and operated above threshold. In contrast to \cite{Li2010} we used filter cavities to generate local oscillators (LO) for a full balanced homodyne (BHD) measurement.

\section{Experimental description and results}																							
\begin{figure}
\includegraphics[angle=-90, width=\columnwidth]{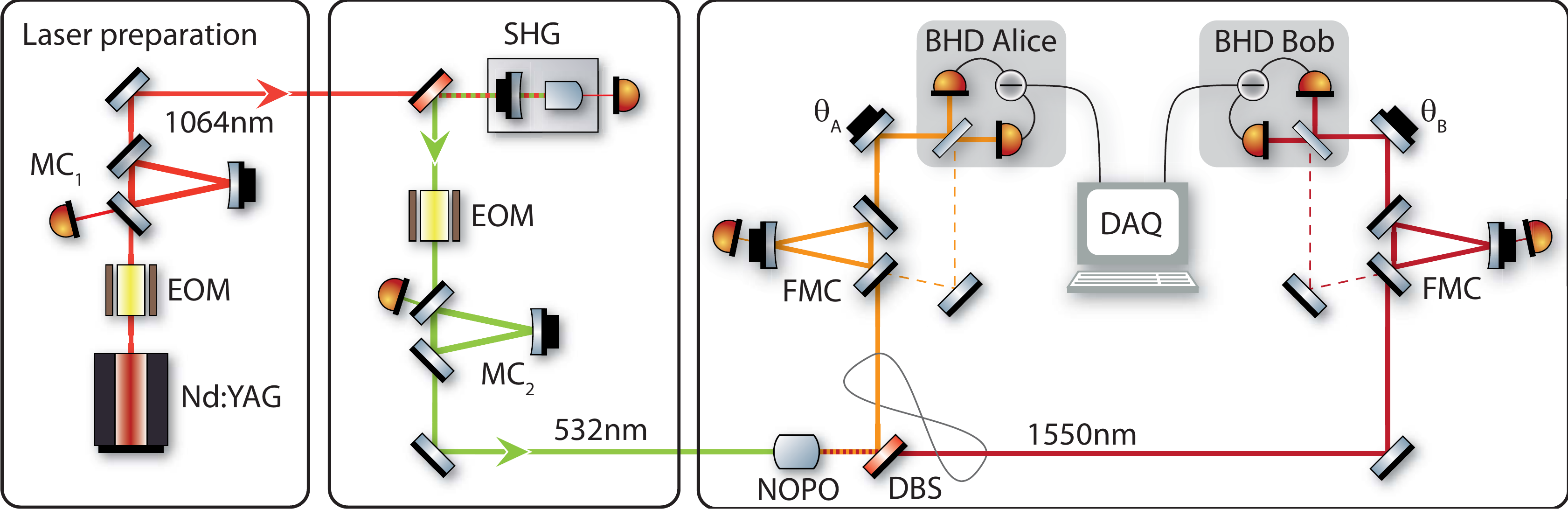}
\caption{Schematic of the setup. After being sent through a mode cleaning cavity (MC\_1) the laser field is frequency doubled in a SHG cavity and filtered by a second mode cleaner (MC\_2) to provide the 532\,nm pump field for the NOPO. The generated twin beams are sent to filter cavities (FMC) in order to spatially separate the sideband and carrier fields. The latter are transmitted through the FMCs and serve as local oscillators (LO) for the balanced homodyne detection (BHD). The reflected sidebands of the twin beams are probed. A data acquisition system is used to record and analyze the data. NOPO, non-degenerate optical parametric oscillator; DBS, dichroic beam splitter; EOM, electro-optical modulator.
} \label{fig:setup}
\end{figure}
\Fig{fig:setup} shows the experimental setup. The light source of our experiment was a Nd:YAG laser of 2.1\,W output power at 1064\,nm. The laser beam was first sent through a ring cavity (MC\_1) with a finesse of $\mf{F}=260$, corresponding to a linewidth of 2.7\,MHz. Reduction of mode distortions of the laser's TEM\_{00} spatial mode profile and technical noise above the cavity's linewidth were ensured. The cavity length was controlled using the Pound-Drever-Hall (PDH) locking scheme \cite{Black01} with a phase modulation at a sideband frequency of 15\,MHz. The output of 1.6\,W was sent directly to the second harmonic generation (SHG) to provide the pump field for the NOPO. 

The SHG was made of a 7\% doped MgO:LiNbO\_3 crystal. The curved back surface of the crystal had a high-reflection coating ($\mathrm{R}=99.96$\%) whereas the flat surface had an anti-reflection coating ($\mathrm{R}<0.05$\%) for both wavelengths. 
The SHG used an out-coupling mirror with power reflectivities of $\mathrm{R}\_{1064\,nm}=90\%$ and $\mathrm{R}\_{532\,nm}<4\%$. The modulation sidebands transmitted through MC\_1 at 15\,MHz were used to control the cavity length with the PDH locking scheme. The generated second-harmonic field had a power of up to 1\,W.

A second filter cavity (MC\_2) was used to ensure a TEM\_{00} spatial mode profile and suppressed technical noise of the pump beam. The finesse was $\mf{F}=560$, corresponding to a linewidth of 1.3\,MHz. A phase modulation at 1.36\,MHz was provided to generate the error signal for MC\_2 and later used cavities, again using the PDH locking scheme.

In our experiment the entanglement was generated in a monolithic standing wave non-linear cavity. The non-linear medium inside the cavity was a periodically poled potassium titanyl phosphate (PPKTP) crystal. 
The phase matching for 532, 810 and 1550\,nm was given at a temperature of 68$^\circ$C, which was stabilized actively. The length of the crystal was 8.9\,mm and the coatings were chosen to form a cavity with a finesse of $\mf{F}=100$ for the twin beams. Hence, the linewidth of both modes was 91\,MHz. The radii of curvature of 8\,mm led to a waist size of 24\,$\mu$m for the pump beam, which simply double passed the crystal. The threshold power was about 120\,mW. The bright output fields were co-propagating and spatially separated with a dichroic beam splitter (DBS). 

\begin{figure}
\includegraphics[width=\columnwidth]{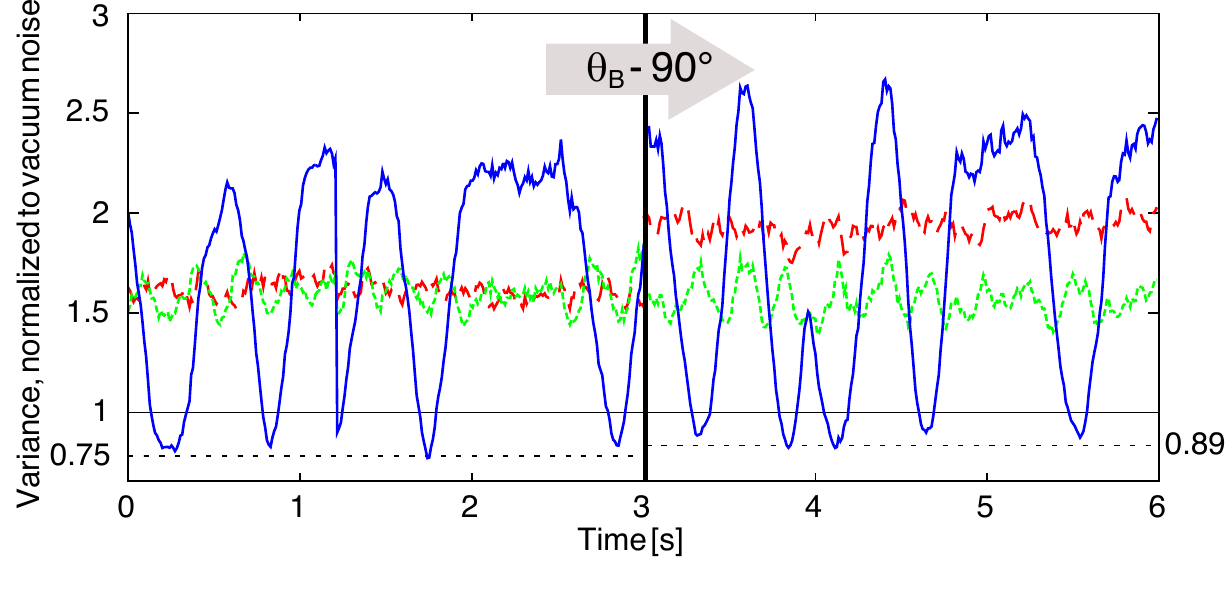}
\caption{Demonstration of entanglement between the two laser fields at Alice's and Bob's sites. Alice's BHD phase is repeatedly ramped up and down and Bob's phase is switched from $\hat{X}$ (left) to $\hat{X}^\perp$ (right). The dotted green and dashed red traces correspond to the variances measured on the individual beams at Alice's and Bob's site, respectively. The minima of the two blue traces correspond to half the variances in Eq.\,(\ref{eq:insep}) thereby fulfilling the inequality with $\mathcal{I}=0.82<1$.} 
\label{fig:result}
\end{figure}

To measure the entanglement by means of BHD detection filter mode cleaner (FMC) were introduced to separate the sideband from the carrier fields \cite{Hage2010}. The latter served as LOs for the BHD measurements. The filter cavities were triangular ring resonators with a finesse of $\mf{F}\approx400$ for 810 and 1550\,nm. 
The twin beams were sent to their corresponding filter cavity and treated in the same manner. An error signal was generated with the PDH scheme using the converted phase modulation of the green pump field at 1.36\,MHz. The transmitted beam served as an optical LO. The reflected part (containing the quantum properties) and the LO were brought to interference on the 50/50 beam splitter. The detection took place with a purpose built photodetector, where the photocurrents were directly subtracted on the circuit board.
Both BHD signals were demodulated at 63.9\,MHz, low pass filtered at 50\,kHz and fed into a data acquisition system. The calculation of the variances of each signal and the variance of the difference of the two signals was conducted by PC software. To ensure a constant LO power, the DC voltage of the FMC were recorded.

To perform the measurement, the NOPO was operated just above threshold and pumped with 130\,mW. We obtained 2.7 and 1.4\,mW LO power, which led to a dark noise clearance of 4 and 6\,dB for the 810 and 1550\,nm detectors, respectively. 

The inseparability criterion \cite{DGCZ00} is a necessary and sufficient criterion for entanglement. 
For our setup it can be written in the form 
\begin{equation}
\mathcal{I}=\frac{1}{4}\left(V(\hat{X}\_A-\hat{X}\_B)+V(\hat{X}\_A^\perp + \hat{X}\_B^\perp)\right) < 1\,. \label{eq:insep}
\end{equation}

$V$ denotes variances, with the variance of a vacuum field normalized to unity. $\hat{X}\_A$ and $\hat{X}\_B$ are the fields' quadrature phase operators at Alice's and Bob's site for which the variance of their difference $V(\hat{X}\_A-\hat{X}\_B)$ is minimal. $\hat{X}\_A^\perp$ and $\hat{X}\_B^\perp$ are the quadrature phase operators orthogonal to $\hat{X}\_A$ and $\hat{X}\_
B$, respectively.
\Fig{fig:result} presents consecutive measurement time series of $V(\hat{X}\_A)$ (dotted green), 
$V(\hat{X}\_B)$ (dashed red), and $1/2\cdot V(\hat{X}\_A-\hat{X}\_B)$ (blue) on the left side and $V(\hat{X}\_A^\perp)$ (dotted green), $V(-\hat{X}\_B^\perp)$ (dashed red), and $1/2\cdot V(\hat{X}\_A^\perp+\hat{X}\_B^\perp)$ (blue) on the right side. During the measurement time shown, the BHD phase at Alice's site was repeatedly ramped up and down and Bob's site was switched from a $\hat{X}$ (left) to a $\hat{X}^\perp$ (right) measurement. Additionally, the vacuum noise levels of the detectors were measured and used to normalize the traces shown. Contemplating the individual variances one recognizes a non-uniform noise distribution among the quadratures. Using Eq.~(\ref{eq:insep}), the data in Fig.~\ref{fig:result} demonstrates the presence of entanglement. The minima of the blue trace on the left side corresponds to a measurement where Alice and Bob were set to their $\hat{X}$ quadrature (green and red trace, lower noise) with $1/2\cdot V(\hat{X}\_A-\hat{X}\_B)=0.75$. On the right side the minima of the blue trace corresponds to a measurement where Alice and Bob were set to their $\hat{X}^\perp$ quadrature (green and red trace, higher noise) with $1/2\cdot V(\hat{X}\_A^\perp+\hat{X}\_B^\perp)=0.89$. Both orthogonal quadratures are showing quantum correlations and together this yields $\mathcal{I}=\frac{1}{2}(0.75+0.89)=0.82<1$, corresponding to $-0.86$\,dB non-classical quadrature correlations. The data presented has not been corrected for the contribution of electronic dark noise. 

\section{Summary}
In conclusion, two-color continuous variable entanglement between the wavelength of 810 and 1550\,nm was experimentally demonstrated. The source was an above threshold driven NOPO 
made of a PPKTP crystal. The entanglement was verified by the inseparability criterion and resulted in $\mf{I}=0.82<1$. By using filter cavities to generate LO for the BHD measurements, we could observe all quadrature phase angles without restrictions to certain frequency bands.


\begin{theacknowledgments}
We would like acknowledge the German Research Foundation, the Centre for Quantum Engineering and Space-Time Research QUEST and the International Max Planck Research School (IMPRS) on Gravitational Wave Astronomy for financial support. 
\end{theacknowledgments}


\end{document}